\documentclass[preprint,11pt]{elsarticle}

\usepackage{graphicx}
\usepackage{dcolumn}
\usepackage{bm}
\usepackage{tipa}
\usepackage{float}
\usepackage{epsf}
\usepackage{epstopdf}
\usepackage{hyperref}
\usepackage{amsfonts,epsfig}
\usepackage{amssymb}
\usepackage{amsmath}  
\usepackage{pdfsync}
\usepackage{slashed}

\usepackage{xcolor}

\newcommand{\be}{\begin{equation}}
\newcommand{\ee}{\end{equation}\noindent}
\newcommand{\bear}{\begin{eqnarray}}
\newcommand{\ear}{\end{eqnarray}\noindent}
\newcommand{\no}{\noindent}
\newcommand{\non}{\nonumber\\}

\definecolor{mbcol}{rgb}{1,0,1}

\usepackage[normalem]{ulem}  

\def\veps#1{\varepsilon_{#1}}
\def\ddel{{}^\bullet\! \Delta}

\def\garg{(\tau,\tau')}

\def\deld{\Delta^{\hskip -.5mm \bullet}}

\def\ddeld{{}^{\bullet}\! \Delta^{\hskip -.5mm \bullet}}

\def\t#1{\tau_1}
\def\mn{{\mu\nu}}

\def\e{\,{\rm e}}

\def\sdel{{\underset{\smile}{\Delta}}}
\def\bcdel{{}^\circ\! {\underset{\smile}{\Delta}}}
\def\bddel{{}^\bullet\! {\underset{\smile}{\Delta}}}
\def\bdeld{{\underset{\smile}{\Delta}^{\hskip -.5mm \bullet}}}

\def\bddeld{{}^{\bullet}\! {\underset{\smile}{\Delta}^{\hskip -.5mm \bullet}}}

\def\bcdeld{{}^{\circ}\! {\underset{\smile}{\Delta}^{\hskip -.5mm \bullet}}}
\def\bccdel{{}^{\circ}\! {\underset{\smile}{\Delta}^{\hskip -.5mm \circ}}}

\def\half{\frac{1}{2}}

\def\kinb{{1\over 4}\dot x^2}

\def\4piTD{{(4\pi T)}^{-{D\over 2}}}
\def\4piT4{{(4\pi T)}^{-2}}

\def\Tintm4{{\dps\int_{0}^{\infty}}{dT\over T}\,e^{-m^2T}
    {(4\pi T)}^{-2}}
\def\Tintm{{\dps\int_{0}^{\infty}}{dT\over T}\,e^{-m^2T}}

\def\t#1{\tau_{#1}}

\newcommand{\slG}{{{\dot G}\!\!\!\! \raise.15ex\hbox {/}}}

\def\GBd12{{\dot G}_{B12}}

\newcommand{\dps}{\displaystyle}

\usepackage[T1]{fontenc} 

\def\half{{1\over 2}}

\def\fourth{{1\over4}}

\def\Eins{{\mathchoice {\rm 1\mskip-4mu l} {\rm 1\mskip-4mu l}
{\rm 1\mskip-4.5mu l} {\rm 1\mskip-5mu l}}}
\def\Z{{\mathchoice {\hbox{$\sf\textstyle Z\kern-0.4em Z$}}
{\hbox{$\sf\textstyle Z\kern-0.4em Z$}}
{\hbox{$\sf\scriptstyle Z\kern-0.3em Z$}}
{\hbox{$\sf\scriptscriptstyle Z\kern-0.2em Z$}}}}
\def\abs#1{\left| #1\right|}

        \def\slash#1{#1\!\!\!\raise.15ex\hbox {/}}
\newcommand{\slD}{\,\raise.15ex\hbox{$/$}\kern-.27em\hbox{$\!\!\!D$}}
\newcommand{\slpartial}{\raise.15ex\hbox{$/$}\kern-.57em\hbox{$\partial$}}

\def\ddel{{}^\bullet\! \Delta}
\def\deld{\Delta^{\hskip -.5mm \bullet}}

\def\ddeld{{}^{\bullet}\! \Delta^{\hskip -.5mm \bullet}}

\def\no{\noindent}

\def\kinb{{1\over 4}\dot x^2}

\def\be{\begin{equation}}
\def\ee{\end{equation}\noindent}
\def\bear{\begin{eqnarray}}
\def\ear{\end{eqnarray}\noindent}
\def\bec{\blue\begin{equation}}
\def\eec{\end{equation}\black\noindent}
\def\bearc{\blue\begin{eqnarray}}
\def\earc{\end{eqnarray}\black\noindent}
\def\benn{\begin{enumerate}}
\def\enn{\end{enumerate}}
\def\ee{&=&}

\def\mn{{\mu\nu}}

\def\e{\,{\rm e}}

\def\b0{{\bf 0}}

\def\4piTD{{(4\pi T)}^{-{D\over 2}}}
\def\4piT4{{(4\pi T)}^{-2}}

\def\Tintm{{\dps\int_{0}^{\infty}}{dT\over T}\,e^{-m^2T}}

\def\cZ{{\cal Z}}

\def\veps{\varepsilon}

\journal{Nuclear Physics B}

\begin{document}
\begin{frontmatter}

\title{Master formulas for the dressed scalar propagator in a constant field}



\author{Aftab Ahmad$^{1,2}$, Naser Ahmadiniaz$^{3,5}$, Olindo Corradini$^{4,6}$, Sang Pyo Kim$^{3,5}$, Christian Schubert$^2$}
\address{$^1$Department of Physics, Gomal University, 29220 D.I. Khan, K.P.K., Pakistan.\\
$^2$Instituto de F\'{\i}sica y Matem\'aticas,
Universidad Michoacana de
San Nicol\'as de Hidalgo. Edificio C-3, Ciudad Universitaria, Morelia 58040, Michoac\'an, M\'exico.\\
$^3$  Center for Relativistic Laser Science, Institute for Basic Science,
Gwangju 61005, Korea\\
$^4$ Dipartimento di Scienze Fisiche, Informatiche e Matematiche, Universit\'a di Modena e Reggio Emilia, Via Campi 213/A, I-41125 Modena, Italy\\
$^5$ Department of Physics, Kunsan National University, Kunsan 54150, Korea\\
$^6$ INFN, Sezione di Bologna, Via Irnerio 46, I-40126 Bologna, Italy}





\begin{abstract}

The worldline formalism has previously been used for deriving compact master formulas for the one-loop 
N-photon amplitudes in both scalar and spinor QED, and in the vacuum as well as in a constant external field. For scalar QED,
there is also an analogous master formula for the propagator dressed with N photons in the vacuum. Here, we extend
this master formula to include a constant field. The two-photon case is worked out explicitly, yielding an integral representation
for the Compton scattering cross section in the field suitable for numerical integration in the full range of electric and magnetic field strengths.

\end{abstract}


\end{frontmatter}


\section{Introduction}
\label{sec:intro}

The one-loop effective action in scalar QED has the well-known ``worldline'' or ``Feynman-Schwinger''
representation \cite{feynman1950},

\bear
\Gamma [A] &=&
-\int_0^{\infty}{dT\over T}\,{\rm e}^{-m^2T}
{\displaystyle \int}_P D x(\tau)
\, \e^{-\int_0^T d\tau 
[ \fourth \dot x^2 + ie \dot x^{\mu}A_{\mu}(x(\tau)) ]}\,.
\label{Gammascal}
\ear
Here $m$ and $T$ denote the mass and proper-time of the loop scalar,
and $\int_{P}D x(\tau)$ the path integral over 
closed loops in (euclidean) spacetime with periodicity $T$ in the proper-time.

Strassler in 1992 \cite{strassler1} showed how to convert this path integral into the following
master formula for the N-photon amplitudes:

\begin{eqnarray}
\Gamma (k_1,\varepsilon_1;\ldots;k_N,\varepsilon_N)
&=&
-{(-ie)}^N
{(2\pi )}^D\delta (\sum k_i)
{\dps\int_{0}^{\infty}}{dT\over T}
{(4\pi T)}^{-{D\over 2}}
\e^{-m^2T}
\prod_{i=1}^N \int_0^T 
d\tau_i
\nonumber\\
&&
\!\!\!\!\!\!\!\!\!\!\!\!\!\!
\times
\exp\biggl\lbrace\sum_{i,j=1}^N 
\Bigl\lbrack  \half G_{Bij} k_i\cdot k_j
-i\dot G_{Bij}\varepsilon_i\cdot k_j
+\half\ddot G_{Bij}\varepsilon_i\cdot\varepsilon_j
\Bigr\rbrack\biggr\rbrace
\Big\vert_{\veps_1\veps_2\cdots\veps_N}
\, .
\nonumber\\
\label{scalarqedmaster}
\end{eqnarray}
\no
Here $G_B,\dot G_B,\ddot G_B$ are the ``bosonic'' worldline Green's function
and its first and second derivatives,

\bear
G_B(\tau,\tau') &\equiv& \abs{\tau-\tau'} -{{(\tau-\tau')}^2\over T}\, , \nonumber\\
\dot G_B(\tau,\tau') &=& {\rm sign}(\tau-\tau') - 2 \frac{\tau-\tau'}{T} \, , \nonumber\\
\ddot G_B(\tau,\tau') &=& 2\delta(\tau-\tau') - \frac{2}{T} \, .
\label{defGB}
\ear
Here a `dot' always means a derivative with respect to the first variable, and 
we abbreviate $G_B(\tau_i,\tau_j) \equiv G_{Bij}$ etc.
$G_B(\tau,\tau')$ is the Green's function for the second derivative operator $\frac{d^2}{d\tau^2}$
adapted to the periodicity, as well as to the ``string-inspired'' (`SI') boundary conditions 

\bear
\int_0^T d\tau \, G_B(\tau,\tau') = \int_0^T d\tau'  \, G_B(\tau,\tau') = 0\,,
\label{SI}
\ear
(up to an irrelevant constant that has been omitted). 
Note that $\ddot G_B(\tau,\tau')$ contains a delta function that brings together two photon legs; this is how the seagull
vertex arises in the worldline formalism. 

The notation $\big\vert_{\veps_1\veps_2\cdots \veps_N}$ means that the exponential should be expanded, and only the
terms linear in each of the polarization vectors be kept. The photons are ingoing and still off-shell, so that these vectors are just book-keeping devices
at this stage.

Originally, the same master formula \eqref{scalarqedmaster} was derived by Bern and Kosower \cite{berkos-prl,berkos-npb} 
from string theory as a generating expression from which to construct the one-loop
on-shell $N$ gluon amplitudes by way of a certain set of rules.   
It contains the information on the $N$ - photon amplitudes
in a form that is not only extremely compact, but also well-organized with respect to 
gauge invariance, particularly when combined with a certain integration-by-parts 
procedure \cite{berkos-prl,berkos-npb,strassler1,91}.  Moreover, it combines into one integral
the various Feynman diagrams differing by the ordering of the $N$ photons. This may not seem very relevant at the
one-loop level, however when the $N$ - photon amplitudes are used as building blocks for multiloop amplitudes it leads to
highly nontrivial representations combining Feynman diagrams of different topologies \cite{15,41} (see also \cite{100}).

In \cite{strassler1} also a generalization to the spinor QED was given (see \cite{41} for 
generalizations to more general field theories). 

Shaisultanov \cite{shaisultanov} then generalized both the scalar and spinor QED master formulas to
the case of QED in a constant external field $F_{\mu\nu}$. For the scalar case, this generalized master formula
can be written as \cite{18,41}

\begin{eqnarray}
&&\Gamma
(k_1,\varepsilon_1;\ldots;k_N,\varepsilon_N)
= -
{(-ie)}^N
{(2\pi )}^D\delta (\sum k_i)
\nonumber\\
&&\hspace{20pt}\times
{\dps\int_{0}^{\infty}}{dT\over T}
{(4\pi T)}^{-{D\over 2}}
\e^{-m^2T}
{\rm det}^{-{1\over 2}}
\biggl[{{\rm sin}({\cal Z})\over {\cal Z}}\biggr]
\prod_{i=1}^N \int_0^T 
d\tau_i
\nonumber\\
&&\hspace{20pt}\times
\exp\biggl\lbrace\sum_{i,j=1}^N 
\Bigl\lbrack \half k_i\cdot {\cal G}_{Bij}\cdot  k_j
-i\varepsilon_i\cdot\dot{\cal G}_{Bij}\cdot k_j
+\half
\varepsilon_i\cdot\ddot {\cal G}_{Bij}\cdot\varepsilon_j
\Bigr\rbrack\biggr\rbrace
\Big\vert_{\veps_1\veps_2\cdots\veps_N}\,,
\nonumber\\
\label{scalarqedmasterF}
\end{eqnarray}
\no
where we have introduced the abbreviation ${\cal Z} \equiv eFT$. 
This master formula differs from the vacuum one, Eq.~\eqref{scalarqedmaster}, only by the additional determinant factor
${\rm det}^{-{1\over 2}}\bigl[{{\rm sin}({\cal Z})\over {\cal Z}}\bigr]$, which represents the dependence of the free   
(photonless) path integral on the external field, and a change of the worldline Green's function $G_B$ to a new
one ${\cal G}_B$ that holds information on the external field,

\bear
{{\cal G}_B}(\tau_i,\tau_j) &=&
{T\over 2{\cal Z}^2}
\biggl({{\cal Z}\over{{\rm sin}({\cal Z})}}
\,{\rm e}^{-i{\cal Z}\dot G_{Bij}}
+ i{\cal Z}\dot G_{Bij} - 1\biggr) \, .
\label{defcalGB}
\ear\no
This Green's function obeys the same SI boundary conditions as the vacuum one, \eqref{SI}. 

The master formula \eqref{scalarqedmasterF} and its spinor QED generalization \cite{shaisultanov,18}
are usually more efficient for the calculation of photonic processes in a  constant field than the standard method
based on Feynman diagrams. Its applications include the vacuum polarization
in a constant field \cite{ditsha,40,41}, photon splitting in a magnetic field \cite{17,41}, and the two-loop Euler-Heisenberg Lagrangian
in an electric/magnetic field~\cite{18,24,korsch,41,66} as well as in a self-dual \cite{51} background field. 
See \cite{baszir1,bacozi2,bacozi1,61,71,76,87} for extensions to gravity and Einstein-Maxwell theory.

Much less has been done for the analogous amplitudes involving an open line. 
For scalar QED in the vacuum, already in 1996 Daikouji et al. \cite{dashsu} obtained the following master formula 
representing the scalar tree-level propagator dressed with $N$ photons (Fig. \ref{fig-multiphoton}):

\begin{figure}[h]
  \centering
   \includegraphics[width=0.7\textwidth]{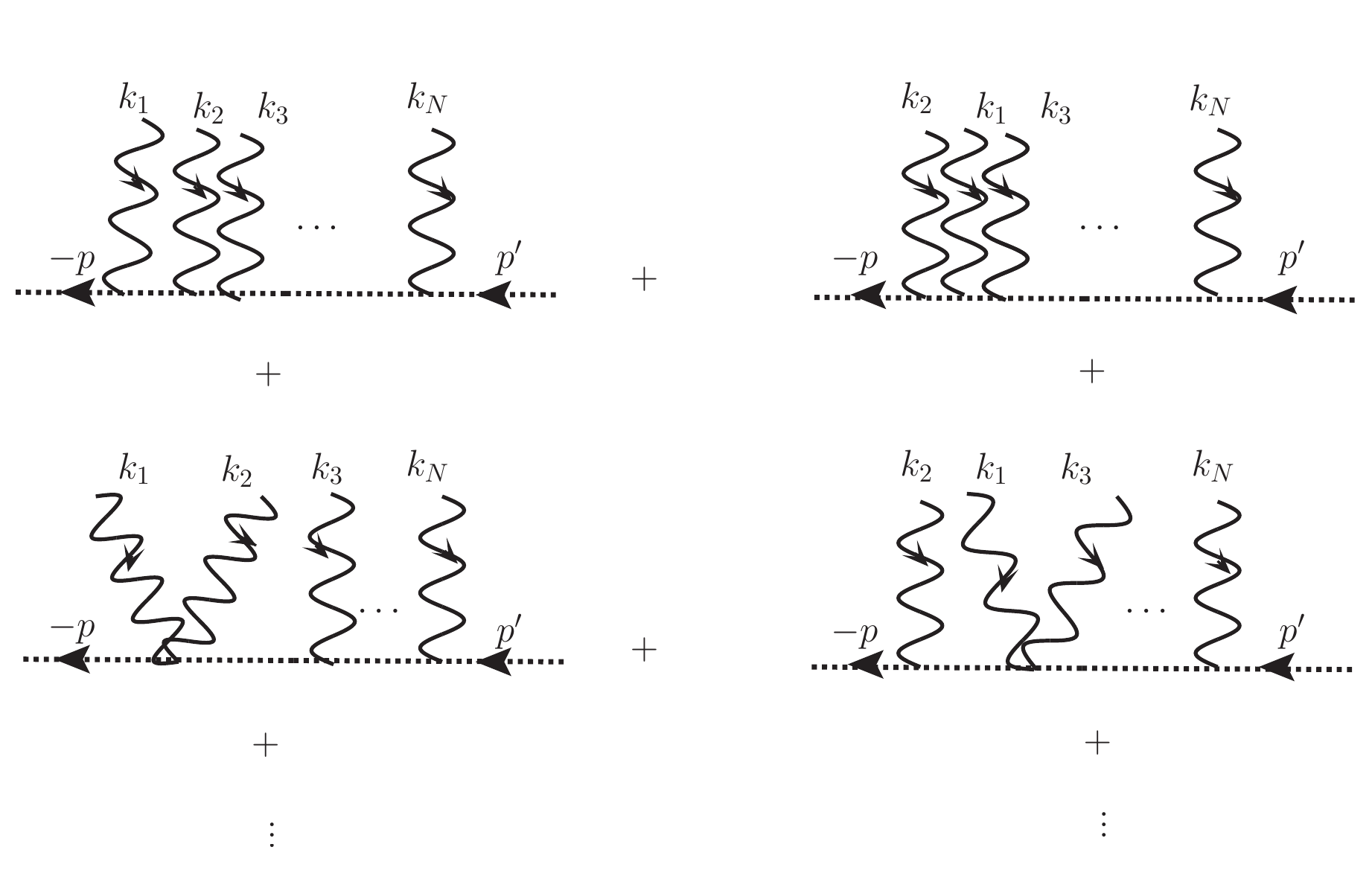}
\caption{Multi-photon Compton-scattering diagram.} 
\label{fig-multiphoton}
\end{figure}

\bear 
D^{pp'}(k_1,\varepsilon_1;\cdots; k_N,\varepsilon_N)&=&(-ie)^N(2\pi)^D\delta^D\Big(p+p'+\sum_{i=1}^Nk_i\Big)\int_0^\infty dT\,{\rm e}^{-m^2T}\nonumber\\
&&\hspace{-1.5cm}\times \prod_{i=1}^N\int_0^Td\tau_i\,{\rm e}^{-T[p+\frac{1}{T}\sum_{i=1}^N(k_i\tau_i-i\varepsilon_i)]^2+\sum_{i,j=1}^N[\Delta_{ij}k_i\cdot k_j-2i\ddel_{ij}\varepsilon_i\cdot k_j-\ddeld_{ij}\varepsilon_i\cdot \varepsilon_j]}\Big\vert_{\varepsilon_1\varepsilon_2\cdots\varepsilon_N}\non
\label{linemaster} 
\ear
Here a different worldline Green's function $\Delta(\tau,\tau')$ appears, 

\bear
\Delta(\tau,\tau')&=&\frac{\tau\tau'}{T}+\frac{\vert\tau-\tau'\vert}{2}-\frac{\tau+\tau'}{2}\,.
\label{defDelta} 
\ear 
Instead of the SI boundary conditions \eqref{SI}, it is adapted to Dirichlet boundary conditions (`DBC')

\bear
\Delta(0,\tau') = \Delta(T,\tau') = \Delta(\tau,0) = \Delta(\tau,T) = 0 \, .
\label{DB}
\ear
Contrary to the former, these boundary conditions break the translation invariance in proper-time,
so that one now has to distinguish between derivatives with respect to the first and the second argument.
A convenient notation is \cite{basvan-book} to use left and right dots to indicate derivatives
with respect to the first and the second argument, respectively:

\bear \ddel(\tau_1,\tau_2)&=&\frac{\tau_2}{T}+\frac{1}{2}{\rm
sign}(\tau_1-\tau_2)-\frac{1}{2}\,,\nonumber\\
\deld(\tau_1,\tau_2)&=&\frac{\tau_1}{T}-\frac{1}{2}{\rm
sign}(\tau_1-\tau_2)-\frac{1}{2}\,,\nonumber\\
\ddeld(\tau_1,\tau_2)&=&\frac{1}{T}-\delta(\tau_1-\tau_2)\,.
\nonumber\\
\label{derDelta}
\ear
The Green's functions $G_B$ and $\Delta$ are related by \cite{6}

\bear
G_B(\tau,\tau') = 2\Delta(\tau,\tau') -  \Delta(\tau,\tau) - \Delta(\tau',\tau') \,,
\label{relgreen}
\ear
(the factor of 2 is conventional) with the inverse relation

\bear
2\Delta(\tau,\tau') = G_B(\tau,\tau') - G_B(\tau,0) - G_B(0,\tau')   \, .
\label{invrelgreen}
\ear
In \cite{dashsu} the master formula \eqref{linemaster} was derived by a comparison with the standard Schwinger parameter integral
representations of the corresponding Feynman diagrams. Recently, the same formula has been rederived \cite{102} inside the worldline formalism,
starting from the generalization of the path integral representation \eqref{Gammascal} to the
propagator of a scalar particle in the Maxwell background:

\bear
D^{xx'}[A] &=&
\int_0^{\infty}
dT\,
\e^{-m^2T}
\int_{x(0)=x'}^{x(T)=x}
Dx\,
\e^{-\int_0^T d\tau\bigl[
\kinb
+ie\,\dot x\cdot A(x)
\bigr]}
\, .
\nonumber\\
\label{scalpropA}
\ear\no
The master formula \eqref{linemaster} so far has been generalized neither to spinor QED,
nor to the inclusion of an external field. 
The purpose of the present paper is to carry out the latter generalization; the extension to the fermonic case
(but without an external field yet) will be presented in a companion paper \cite{111}. See~\cite{ahbaco-colorful} for
a non-abelian generalization of the dressed scalar propagator. 

The organization of the paper is as follows.
As a warm-up, in section \ref{nophot} we use the path integral representation to rederive the well-known scalar propagator in
a constant field, in configuration as well as in momentum space. 
In section \ref{dressed} we obtain our master formulas for the photon-dressed propagator in a constant field in both configuration
and momentum space, generalizing the vacuum calculation of \cite{102}.
In section \ref{compton} we work the momentum space formula out for the $N=2$ case, and obtain
a compact integral representation for the Compton scattering cross section in a constant field.  
Section~\ref{summary} provides a summary and outlook.
In appendix \ref{app-conv} we give our conventions, while in appendix \ref{app-green} we collect  
some information on the constant field worldline Green's functions. 

\section{The propagator in a constant field}
\label{nophot}

In this section, we use \eqref{scalpropA} to just rederive the well-known scalar propagator in a constant field, without photons yet.

\subsection{Configuration space}

Choosing Fock-Schwinger gauge, the gauge potential for a constant field can be written as

\begin{eqnarray}
A^{\mu}(y)=-\frac{1}{2}F^{\mu\nu}(y-x^\prime)^{\nu},\label{FS}
\end{eqnarray}
where we have fixed the initial point of the trajectory $x'$ as the reference point where the potential will vanish.

Further, we decompose the arbitrary trajectory $x(\tau)$ into a straight-line part and a fluctuation part $q(\tau)$
obeying Dirichlet boundary conditions, $q(0) = q(T)=0$:

\begin{eqnarray}
x(\tau)=x^{\prime} +\frac{\tau}{T}(x-x^\prime)+ q(\tau).\label{split}
\end{eqnarray}
Using \eqref{FS} and \eqref{split} in \eqref{scalpropA}, and further defining

\bear
Q^{\mu} \equiv \int_0^T d\tau q^{\mu}(\tau) \, ,
\label{defQ}
\ear
after some simple manipulations the integrand can be rewritten as

\begin{eqnarray}
D^{xx'}(F) &=&
\int^{\infty}_{0}dT \e^{-m^2T} \e^{-\frac{(x-x^{\prime})^2}{4T}} \int Dq(\tau) \e^{-\int^{T}_{0}d\tau  \frac{\dot{q}^2}{4}+\frac{ie}{2} \int^{T}_{0} d\tau  \dot{q}^\mu F_{\mu\nu} q^{\nu}+\frac{ie}{T}(x-x^{\prime})^\mu F_{\mu\nu} Q^{\nu}}
\nonumber\\
&=& 
\int^{\infty}_{0}dT \e^{-m^2T} \e^{-\frac{(x-x^{\prime})^2}{4T}} \int Dq(\tau) \e^{-\int^{T}_{0}d\tau \frac{1}{4}q\left(-\frac{d^2}{d\tau^2}+2ieF\frac{d}{d\tau} \right)q+  \frac{ie}{T}(x-x^{\prime}) F Q}
\, .
\nonumber\\\label{20}
\end{eqnarray}
The path integral is already of gaussian form, and in the second line we have written it in a form that prepares the formal gaussian integration.
Apart from the free path integral normalization, which is (see, e.g., \cite{103})

\begin{eqnarray}
\int Dq(\tau) \e^{- \int^{T}_{0} d\tau q \left(-\frac{1}{4}\frac{d^{2}}{d\tau^{2}}\right)q}=(4\pi T)^{-\frac{D}{2}} \, ,\label{7}
\end{eqnarray}
this involves the determinant and the inverse of the operator $-\frac{d^2}{d\tau^2}+2ieF\frac{d}{d\tau}$.  
For the case of the SI boundary conditions \eqref{SI}, the relevant formulas have been given already in \eqref{scalarqedmasterF}, \eqref{defcalGB} above.
The ratio of the field-dependent and free path integral normalizations are

\bear
\frac
{ 
{\rm Det'}^{-\half}_P\bigl(-\fourth \frac{d^2}{d\tau^2}+ \half ie F \frac{d}{d\tau} \bigr)
}
{
{\rm Det'}^{-\half}_P\bigl(-\fourth \frac{d^2}{d\tau^2} \bigr)
}
=
{\rm Det'}^{-\half}_P\Bigl(\Eins -2 ie F \bigl(\frac{d}{d\tau}\bigr)^{-1} \Bigr)
=
{\rm det}^{-{1\over 2}}\Bigl[{{\rm sin}({\cal Z})\over {\cal Z}}\Bigr]\,,
\label{detratio}
\ear
(the `prime' refers to the elimination of the zero mode which is contained in the path integral for string-inspired boundary conditions).
This can be shown by a direct eigenvalue computation \cite{103}, and it is easy to see that the spectrum does not change when passing
from string-inspired to Dirichlet boundary conditions, so that \eqref{detratio} holds unchanged for the open-line case. 

The worldline Green's function does change, but still relates to the one for string-inspired boundary conditions in the same way as in the
vacuum case, \eqref{invrelgreen}:

\begin{eqnarray}
\sdel(\tau,\tau^{\prime})
\equiv \langle \tau\mid \Bigl( \frac{d^2}{d\tau^2} - 2ieF\frac{d}{d\tau}\Bigr)^{-1}\mid \tau'\rangle_{\rm DBC} 
=\half \Bigl( {\cal G}_{B}(\tau,\tau^{\prime})-{\cal G}_{B}(\tau, 0)-{\cal G}_{B}(0,\tau^{\prime})+{\cal G}_{B}(0,0)\Bigr)
\, .
\nonumber\\
\label{defDeltafriendly} 
\end{eqnarray}
Note that, contrary to the vacuum Green's function \eqref{defDelta}, it is a non-trivial matrix in the Lorentz space-time indices. 
Using this Green's function in the usual completing-the-square procedure, we get  

\begin{eqnarray}
D^{xx'}(F) &=&
\int^{\infty}_{0}dT \e^{-m^2T} \e^{-\frac{(x-x^{\prime})^2}{4T}}
(4\pi T)^{-\frac{D}{2}} {\rm det}^{-\half}\left[\frac{\sin(eFT)}{eFT} \right] \nonumber\\ 
&& \times {\rm exp}\biggl\lbrace \int^{T}_{0} d\tau \int^{T}_{0} d\tau^{\prime} \frac{ie}{T} (x-x^\prime) F \sdel(\tau,\tau^{\prime})\frac{ie}{T}F(x-x^\prime) \biggr\rbrace 
\nonumber\\
&=&  
\int^{\infty}_{0}dT \e^{-m^2T} \e^{-\frac{(x-x^{\prime})^2}{4T}}
(4\pi T)^{-\frac{D}{2}} {\rm det}^{\half}\left[\frac{\cal Z}{\sin \cal Z} \right] \nonumber\\ 
&& \times {\rm exp}\biggl\lbrace - \frac{1}{T^4} (x-x^\prime) {\cal Z}\bccdel{\cal Z}(x-x^\prime) \biggr\rbrace 
\, .
\nonumber\\
\label{22}
\end{eqnarray}
Here we have now extended the above `dot' notation to include integration as well as differentiation; a left (right) `open circle' on $\sdel(\tau,\tau')$ denotes
an integral $\int_0^T d\tau$ ($\int_0^Td\tau'$). 
In appendix \ref{app-green} we show that 

\bear
\bccdel \equiv \int_0^T d\tau \int_0^Td\tau' \sdel (\tau,\tau') = \frac{T^3}{4{\cZ}} \Bigl(\cot \cZ- \frac{1}{\cZ}\Bigr)
\, , 
\label{int}
\ear
which brings us to the well-known proper-time representation of the constant-field propagator (see, e.g., \cite{frgish-book}),

\begin{eqnarray}
D^{xx'}(F) &=&
\int^{\infty}_{0}dT \e^{-m^2T} 
(4\pi T)^{-\frac{D}{2}} {\rm det}^{\half}\left[\frac{\cal Z}{\sin \cal Z} \right]  {\rm exp}\biggl\lbrace - \frac{1}{4T} (x-x^\prime)\cZ\cot \cZ (x-x^\prime) \biggr\rbrace 
\, .
\nonumber\\
\label{xpropF}
\end{eqnarray}

\subsection{Momentum space}

We Fourier transform (\ref{xpropF}),

\begin{eqnarray}
D^{pp^{\prime}}(F) =\int d^{D}x \int d^{D}x^\prime \e^{ip\cdot x+ip^{\prime}\cdot x^{\prime}}D^{xx'}(F)\,, 
\label{fourier}
\end{eqnarray}
and changing the integration variables to 

\bear
x_+ = \half (x+x^{\prime}), \quad x_- =  x-x^{\prime}\,,
\label{defxpm}
\ear
we get

\begin{eqnarray}
D^{pp^{\prime}}(F)  &=& \int^{\infty}_{0}dT \e^{-m^2T} 
(4\pi T)^{-\frac{D}{2}} {\rm det}^{\half}\left[\frac{\cal Z}{\sin \cal Z} \right]  \nonumber\\
&&\!\!\!\!\!\!\!\times 
\int d^{D}x_{-} \int d^{D}x_{+}\, \e^{\frac{i}{2}(p - p^{\prime})\cdot x_{-} +i(p+p^\prime) x_{+}}
 {\rm exp}\biggl\lbrace - \frac{1}{4T} x_-\cZ\cot \cZ x_- \biggr\rbrace 
 \, .\non
\end{eqnarray}
As

\begin{eqnarray}
\int d^{D}x_{-}  \e^{ip\cdot x_{-} - \frac{1}{4T} x_-\cZ\cot \cZ x_-}
= \frac{(4\pi T)^{\frac{D}{2}}}{{\rm det}^{\half}\left[\cZ \cot\cZ\right]}
\e^{-T p\left(\cZ{\cot\cZ}\right)^{-1} p }\,,
 \label{27}
\end{eqnarray}
the final result becomes

\begin{eqnarray}
D^{pp^{\prime}}(F) 
&=&(2\pi)^{D} \delta(p + p^{\prime}) D(p,F)\,, \nonumber\\
D(p,F) &=& 
\int^{\infty}_{0}dT \, \e^{-m^2T}\, \frac{\e^{-T p (\frac{{\rm tan\cZ}}{\cZ}) p}}{{\det}^{\half}\left[{\rm cos\cZ} \right]}\, . \nonumber\\\label{28}
\end{eqnarray}

\section{The dressed propagator in a constant field}
\label{dressed}

We now wish to dress the propagator with $N$ photons in addition to the constant field. 
As before, we start in configuration space. 

\subsection{Configuration space}
\label{dressed-xspace}

For this purpose, the potential in \eqref{scalpropA} has to be chosen as 

\bear
A = A_{\rm ext} + A_{\rm phot}\,,
\label{Atotal}
\ear
where $A_{\rm ext}$ is the same as in \eqref{FS}, and $A_{\rm phot}$ represents a sum of 
plane waves:

\bear
A_{\rm phot}^{\mu}(x) = \sum_{i=1}^N \varepsilon^{\mu}_i\,\e^{ik_i \cdot x} \, .
\label{Apw}
\ear
Each photon then effectively gets represented by a vertex operator 

\bear 
V^A[k,\varepsilon]=\int_0^Td\tau\, \veps\cdot\dot{x}(\tau)\,{\rm e}^{ik\cdot x (\tau)}\,,
\label{vertop}
\ear
integrated along the scalar line. This 
leads to the following path integral representation of the 
constant-field propagator dressed with $N$ photons:

\bear
D^{xx'}(F\mid k_1,\veps_1;\cdots;k_N,\veps_N) &=& 
(-ie)^N 
\int_0^{\infty}
dT\,
\e^{-m^2T}
\int_P
Dx\,
\e^{-\int_0^T d\tau\bigl[
\kinb
+ie\,\dot x\cdot A_{\rm ext}(x)
\bigr]}
\nonumber\\
&& \times
V[k_1,\varepsilon_1] V[k_2,\varepsilon_2]\cdots V[k_N,\varepsilon_N]
\, .
 \nonumber\\
\label{DNpointscal}
\ear
For the evaluation of the path integral, it will be convenient to rewrite the photon vertex operator \eqref{vertop} as

\bear 
V^A[k,\varepsilon] =\int_0^Td\tau\, {\rm e}^{ik\cdot x(\tau)+\veps\cdot \dot{x}(\tau)}\Big\vert_{{\rm lin}(\veps)}
\,.
\label{vertopexp}
\ear
Applying the path decomposition \eqref{split} we get the following generalization of \eqref{20},

\bear
D^{xx'}(F\mid k_1,\veps_1;\cdots;k_N,\veps_N) &=& 
(-ie)^N\int_0^{\infty}dT {\rm e}^{-m^2T-\frac{x_-^2}{4T}}\int Dq
\, \e^{-\int^{T}_{0}d\tau \frac{1}{4}q\left(-\frac{d^2}{d\tau^2}+2ieF\frac{d}{d\tau} \right)q+  \frac{ie}{T}x_- F Q}
\nonumber\\
&\times&\int_0^T\prod_{i=1}^Nd\tau_i\,
\,{\rm e}^{\sum_{i=1}^N\big(\veps_i\cdot \frac{x_-}{T}+\veps_i\cdot\dot q(\tau_i)+ik_i\cdot x_-\frac{\tau_i}{T} +ik_i\cdot x'+ik_i\cdot q(\tau_i)\big)}
\Big\vert_{\veps_1\veps_2\cdots \veps_N}\,. 
\nonumber\\
\label{master-open-scalar}
\ear
The path integral is already in a form suitable for gaussian integration. ``Completing the square'' using the Green's function \eqref{defDeltafriendly}, and 
using \eqref{int}, we get the following $x$ - space master formula:

\bear
D^{xx'}(F\mid k_1,\veps_1;\cdots;k_N,\veps_N) &=& 
(-ie)^{N}\int^{\infty}_{0}dT \e^{-m^2T}(4\pi T)^{-\frac{D}{2}}
{\det}^{\half}\left[\frac{\cZ}{\sin\cZ} \right]
{\rm e}^{- \frac{1}{4T} x_- \cZ\cot \cZ x_-}
\nonumber\\ 
&& \hspace{-1pt}\times\int^{T}_{0}d\tau_{1}\cdots \int^{T}_{0}d\tau_{N}
\, \e^{\sum^{N}_{i=1} \bigl(\varepsilon_{i}\cdot\frac{x_{-}}{T}+ik_{i}\cdot\frac{x_{-}\tau_{i}}{T}+ik_{i}\cdot x' \bigr)}
\nonumber\\
&&\hspace{-140pt} \times
{\rm exp}\bigg[  \sum^{N}_{i,j=1}\Bigl( k_{i}~\sdel{}_{ij}~ k_{j}-2i\varepsilon_{i}~\bddel{}_{ij}~ k_{j}-\varepsilon_{i}~\bddeld{}_{ij}~\varepsilon_{j}\Bigr) 
 +\frac{2e}{T}x_- \sum^{N}_{i=1}  \Bigl(F~\bcdel{}_{i}~k_{i}- iF~\bcdeld{}_{i}~\varepsilon_{i}\Bigr)\bigg] \Big\vert_{\veps_1\veps_2\cdots \veps_N}\,. 
\nonumber\\
   \label{master-open-scalar-2}
\ear
For the special case of a purely magnetic field, this $x$ - space master formula was obtained already in 1994 by McKeon and Sherry
\cite{mckshe}.

\subsection{Momentum space}
\label{dressed-pspace}

The transition to momentum space is quite analogous to the photon-less case. 
We Fourier transform according to \eqref{fourier}, and change the variables to \eqref{defxpm}. 
The $x_+$ integral produces the global delta function for energy-momentum conservation, and the $x_-$
integral is gaussian.  Performing it we get our momentum space master formula:

\bear
D^{pp'}(F\mid k_1,\veps_1;\cdots;k_N,\veps_N) &=& 
(-ie)^N (2\pi)^{D} \delta \Bigl( p+ p^\prime+ \sum^{N}_{i=1}k_{i}\Bigr)   \int^{\infty}_{0}dT \e^{-m^2T}
\frac{1}{{\det}^{\half}\left[{\rm cos\cZ} \right]}
\nonumber\\ 
&&\hspace{-90pt}\times\int^{T}_{0}d\tau_{1}\cdots \int^{T}_{0}d\tau_{N}
\, \e^{\sum^{N}_{i,j=1}\bigl( k_{i}\, \sdel{}_{ij}\, k_{j}-2i\varepsilon_{i}\, \bddel{}_{ij}\, k_{j}-\varepsilon_{i}\, \bddeld{}_{ij}\, \varepsilon_{j}\bigr)} 
\e^{-T b (\frac{{\rm tan\cZ}}{\cZ}) b}
\Big\vert_{\veps_1\veps_2\cdots \veps_N}
\, .
\nonumber\\
\label{master-pspace}
\ear
Here we have defined

\begin{eqnarray}
b \equiv p+\frac{1}{T}\sum^{N}_{i=1}\Bigl[\Bigl( 
  \tau_{i}-2ie F~\bcdel{}_{i}\Bigr) k_i -i \left(1-2ie F~\bcdeld{}_{i}  \Bigr)  \varepsilon_{i} \right] \, .
  \label{defb}
\end{eqnarray}

The master formula \eqref{master-pspace} describes the same set of Feynman diagrams depicted in Fig.~\ref{fig-multiphoton}, only that
now all the scalar propagators are the ``full'' ones in the external field (usually indicated by a double line). 
When applying it to the calculation of physical processes, one has to take into account that it describes the {\it untruncated} dressed propagator,
i.e. the final propagators on each end of the scalar line in Fig.~\ref{fig-multiphoton} are included. To obtain the matrix element ${\cal T}$, 
we have to cancel these final propagators using \eqref{28}:

\bear
{\cal T}(F\mid k_1,\veps_1;\cdots;k_N,\veps_N)  &=& 
\frac{D^{pp'}(F\mid k_1,\veps_1;\cdots;k_N,\veps_N)}
{D(p,F)D(p',F)}
\, .
\nonumber\\
\label{calM}
\ear
Moreover, it will be convenient to Wick rotate from euclidean to Minkowski space; the rules for the Wick rotation are given in appendix~\ref{app-conv} together with our conventions. 
In appendix~\ref{app-green} we collect the formulas necessary to write the integrand in explicit form. 
We use~\eqref{defDeltafriendly} to write the Green's function $ \sdel(\tau,\tau^{\prime})$ in terms of 
${\cal G}_B(\tau,\tau^{\prime})$, which is translation invariant and obeys~\eqref{SI} which will be very useful here. 
We then explain how to write ${\cal G}_B(\tau,\tau^{\prime})$ explicitly for a generic constant field. 

Finally, let us remark that eventual poles in the global proper-time integral due to the determinant factor in \eqref{master-pspace} 
are spurious, because when ${\cos \cZ} = 0$ the factor $\e^{-T b (\frac{{\rm tan\cZ}}{\cZ}) b}$ will vanish too (differently from the
corresponding one-loop amplitudes, where such poles lead to an imaginary part related to pair creation).

\section{Compton scattering in a constant field}
\label{compton}

We will now work out the $N=2$ case, i.e. Compton scattering in a constant field. 
Expanding out the exponentials in \eqref{master-pspace} and projecting to the terms linear in both polarization vectors,
we find (omitting now the global factor for energy-momentum conservation):

\bear
D^{pp'}(F\mid k_1,\veps_1;k_2,\veps_2) &=& 
e^2  \int^{\infty}_{0}dT\frac{\e^{-m^2T}}{{\det}^{\half}\left[{\rm cos\cZ} \right]}\nonumber\\
&\times&\int^{T}_{0}d\tau_{1}\int^{T}_{0}d\tau_{2} \e^{-T b_0 (\frac{{\rm tan\cZ}}{\cZ}) b_0+\sum^{2}_{i,j=1} k_{i}\, \sdel{}_{ij}\, k_{j}}
\veps_1 M_{12} \veps_2\,,\non
\label{master-expanded}
\ear
with

\bear
b_0 &\equiv & p+\frac{1}{T}\sum^{2}_{i=1}\Bigl(\tau_{i}-2ie F~\bcdel{}_{i}\Bigr) k_i \,,
\label{defb0}
\ear
and

\bear
M_{12} &\equiv & 2 \bddeld{}_{12}
-\frac{2}{T} \Bigl(1+2ie \bcdeld{}_1^TF\Bigr) \frac{{\rm tan\cZ}}{\cZ}\Bigl(1-2ieF\, \bcdeld{}_2\Bigr)
\nonumber\\&&\hspace{-5pt}
+ 4 \Bigl[(1+2ie\bcdeld{}_1^TF)  \frac{{\rm tan\cZ}}{\cZ}b_0-\sum_{i=1}^2\bddel{}_{1i}~k_i\Bigr]
\Bigl[b_0 \frac{{\rm tan\cZ}}{\cZ}\Bigl(1-2ieF\, \bcdeld{}_2\Bigr) - \sum_{i=1}^2 k_i\bdeld{}_{i2}\Bigr]
\, .
\nonumber\\
\label{M}
\ear
Squaring, and performing the sum over the photon polarizations via 

\bear
\sum_{\rm pol}\veps_{i}^{\ast\mu}\veps_{i}^{\nu} \longrightarrow g^\mn\,,
\label{polsum}
\ear
we get the following for the Compton cross section:

\bear
\sum_{\rm pol}\,{\cal T}^{\ast}{\cal T} 
&=&
\frac{e^4}{\abs{D(p,F)}^2\abs{D(p',F)}^2}
\nonumber\\
&&\times
\int^{\infty}_{0}dT'  \frac{\e^{-m^2T'}}{{\det}^{\half}\left[{\rm cos\cZ'} \right]}
\int^{T'}_{0}d\tau'_{1}\int^{T'}_{0}d\tau'_{2}
\, \e^{-T' b_0^{\ast} (\frac{{\rm tan\cZ'}}{\cZ'}) b_0^{\ast}+\sum^{2}_{i,j=1} k_{i}\, \sdel'{}_{ij}\, k_{j}}
\nonumber\\
&&\times
\int^{\infty}_{0}dT  \frac{\e^{-m^2T}}{{\det}^{\half}\left[{\rm cos\cZ} \right]}
\int^{T}_{0}d\tau_{1}\int^{T}_{0}d\tau_{2}
\, \e^{-T b_0 (\frac{{\rm tan\cZ}}{\cZ}) b_0+\sum^{2}_{i,j=1} k_{i}\, \sdel{}_{ij}\, k_{j}}
\, 
{\rm tr}(M_{12}^{\prime\dag}M_{12})\, .
\nonumber\\
\label{comptonfin}
\ear
After writing the integrand explicitly with the help of the formulas of appendix~\ref{app-green}, this expression is suitable
for numerical integration.

\section{Summary and Outlook}
\label{summary}

Using the worldline path integral formalism, we have derived a Bern-Kosower type master formula for the scalar propagator in QED, in a constant field and  dressed by an arbitrary number
of photons. The $x$ - space version of this formula generalizes the one obtained by McKeon and Sherry for the purely magnetic case \cite{mckshe}; the $p$ - space version
generalizes the vacuum master formula of Daikouji et al. \cite{dashsu} on one hand, the closed-loop master formula of Shaisultanov \cite{shaisultanov} on the other. 
Our master formula is valid off-shell, and combines the various orderings of the $N$ photons along the scalar line. 
It can thus be used as a convenient starting point for the construction of higher-loop scalar QED processes in a constant field.
On-shell, it yields parameter integral representations for linear and nonlinear Compton scattering in the field, as well as the various processes related to it
by crossing.

 To make this paper self-contained, we have also provided all the machinery necessary for writing the integrands in explicit form. 
We have worked out the integrand for the linear Compton scattering case explicitly, arriving at a compact representation suitable for numerical integration.
The results of such a numerical computation will be presented in a forthcoming publication. Compton scattering in magnetic fields is a process of potential relevance for astrophysics, but, to the best of
our knowledge, so far has been studied only in the strong-field limit \cite{daughhard86}.

\section*{Acknowledgments}

We would like to thank Gerry McKeon for discussions. 
A. A. acknowledges Paola Rioseco for the support and motivation. 
The work of N. A. and S. P. K was supported by IBS (Institute for Basic Science) under grant IBS-R012-D1, N. A thanks the Universit\`a di Modena e Reggio Emilia for their warm hospitality while parts of this work were completed. 
C. S. thanks CONACYT for financial support through grant CB14-242461.

\appendix

\section{Conventions}
\label{app-conv}

At the path integral level, we work in the Euclidean space with
a positive definite metric
$(g_{\mu\nu})={\,\rm diag}(++\ldots +)$.
The euclidean field strength tensor is defined by
$F^{ij}= \varepsilon_{ijk}B_k, i,j = 1,2,3$,
$F^{4i}=-iE_i$.
Minkowski space amplitudes are obtained by analytically continuing

\bear
g_{\mu\nu}&\rightarrow& \eta_{\mu\nu} \, ,\non
k^4&\rightarrow& -ik^0 \, ,\non
T&\rightarrow& is \, ,\non
F^{4i}&\rightarrow& F^{0i}=E_i \, .\non
\label{euctomink}
\ear\no
where
$(\eta_{\mu\nu}) = {\,\rm diag}(-+++)$.
These Minkowski space conventions agree with \cite{srednicki-book} up to the sign of the
charge $e$. 

\no
Momenta appearing in vertex operators are {\sl ingoing}.

\section{Worldline Green's functions}
\label{app-green}

Here we collect the information necessary to work out explicitly the integrands generated by the master formulas
\eqref{master-open-scalar-2} and \eqref{master-pspace} for any $N$. 

\subsection{Expressing the DBC Green's function through the SI one}

Rather than writing out the DBC Green's function $\sdel(\tau,\tau')$ 
and its derivatives directly in terms of trigonometric functions of the field strength tensor, we find
it convenient to first rewrite them in terms of the SI Green's function ${\cal G}_B(\tau,\tau')$
via \eqref{defDeltafriendly}, 

\begin{eqnarray}
\sdel(\tau,\tau^{\prime})
=\half \Bigl( {\cal{G}}_{B}(\tau,\tau^{\prime})-{\cal{G}}_{B}(\tau, 0)-{\cal{G}}_{B}(0,\tau^{\prime})+{\cal{G}}_{B}(0,0)\Bigr)
\, .
\nonumber\\
\label{appB-defDeltafriendly} 
\end{eqnarray}
The advantages of ${\cal G}_B(\tau,\tau')$ are that it is translation invariant, so that we do not have to distinguish between right and left
derivatives, and that it fulfills the same nonlocal boundary conditions 
as the vacuum Green's function \eqref{SI},

\bear
\int^{T}_{0}d\tau\,{\cal G}_{B}(\tau,\tau^{\prime})=
\int^{T}_{0}d\tau\, \dot{\cal G}_B(\tau,\tau^{\prime})=
\int^{T}_{0}d\tau\,\ddot{\cal G}_B(\tau,\tau^{\prime})= 0
\, .
\label{app-SI}\ear

The latter property will be very useful for the `circled' Green's functions.  
Moreover, the Lorentz matrix structure of ${\cal G}_B(\tau,\tau')$ has already been worked out for the various types of constant fields \cite{40,41}. 

Using \eqref{appB-defDeltafriendly}, the various derivatives and integrals of 
$\sdel(\tau,\tau')$ appearing in the master formulas become

\bear
 \bddel \garg &=& \half \Bigl( \dot{\cal G}_{B}(\tau,\tau^{\prime})-\dot{\cal G}_{B}(\tau, 0)\Bigr)\,,\nonumber\\
 \bddeld \garg &=&  -  \half \ddot{\cal G}_{B}(\tau,\tau^{\prime})\,, \nonumber\\
 \bcdel (\tau') &=&
\frac{T}{2} \Bigl(-{\cal{G}}_{B}(0,\tau^{\prime})+{\cal{G}}_{B}(0,0)\Bigr)\,,
  \nonumber\\
  \bcdeld (\tau') &=& \frac{T}{2} \dot{\cal{G}}_{B}(0,\tau^{\prime})\,,\nonumber\\
\bccdel &=&
\frac{T^2}{2} {\cal{G}}_{B}(0,0)\,.
\nonumber\\
\label{DeltatoG}
\ear

\subsection{General properties of the Green's function ${\cal G}_B$}

Here we cite a few general properties of the Green's function ${\cal G}_B$ and its
derivatives; for derivations and more details, see \cite{40,41,76}. 
We can write these functions as power series in the matrix  ${\cal Z} \equiv eFT$ as follows \cite{18}:

\begin{eqnarray}
{{\cal G}_B}(\tau,\tau') &=&
{T\over 2{\cal Z}^2}
\biggl({{\cal Z}\over{{\rm sin}({\cal Z})}}
\,{\rm e}^{-i{\cal Z}\dot G_B\garg}
+ i{\cal Z}\dot G_B\garg - 1\biggr) \, ,
\nonumber\\
\dot{\cal G}_B(\tau,\tau')
&=&
{i\over {\cal Z}}\biggl({{\cal Z}\over{{\rm sin}({\cal Z})}}
{\rm e}^{-i{\cal Z}\dot G_B\garg}-1\biggr) \, ,
\nonumber\\
\ddot{\cal G}_{B}(\tau,\tau')
&=& 2\delta (\tau -\tau') -{2\over T}{{\cal Z}\over{{\rm sin}({\cal Z})}}
{\rm e}^{-i{\cal Z}\dot G_B\garg} \, .
\nonumber\\
\label{app-GB}
\end{eqnarray}
\noindent
By absorbing the dependence on $\tau,\tau'$ in terms of the derivative of the vacuum Green's function, $\dot G_B\garg$, one avoids having to make an explicit case distinction
between $\tau_1 >\tau_2$ and $\tau_1 <\tau_2$ that would become necessary otherwise \cite{shaisultanov}. 
Let us note also the coincidence limits of ${\cal G}_B, \dot{\cal G}_B$:

\bear
{\cal G}_{B}(\tau,\tau)&=&
{T\over 2{{\cal Z}}^2}
\Bigl({\cal Z}\cot({\cal Z})-1
\Bigr) \, ,
\nonumber\\
\dot {\cal G}_B(\tau,\tau) &=& i{\rm cot}({\cal Z}) -{i\over {\cal Z}} \,.
\nonumber\\
\label{coincalG}
\end{eqnarray}
Note that they are independent of $\tau$. 
As Lorentz matrices, ${\cal G}_B$ and its derivatives have the following symmetry properties:

\bear
{\cal G}_B(\tau,\tau') = 
{\cal G}_B^{T}(\tau',\tau),
\quad
\dot{\cal G}_B(\tau,\tau') = 
-\dot{\cal G}_B^{T}(\tau',\tau),
\quad
\ddot{\cal G}_B(\tau,\tau') = 
\ddot{\cal G}_B^{T}(\tau',\tau)\, .
\nonumber\\
\label{symmcalGB}
\ear\no
For weak background fields, it is often justified to approximate the Green's function by the first few terms of its expansion in $F_{\mn}$. 
To order $F^2$, one finds

\begin{eqnarray}
{\cal G}_B \garg &=& G_B\garg-{T\over 6}
-{i\over 3}
\dot G_B\garg G_B\garg TeF+\Bigl({T\over 3}G_B^2\garg
-{T^3\over 90}\Bigr){(eF)}^2+O(F^3)\, ,
\nonumber\\
\dot{\cal G}_B\garg
&=&\dot G_B\garg +2i\Bigl(G_B\garg -{T\over 6}\Bigr)eF
+{2\over 3}\dot G_B\garg G_B\garg T{(eF)}^2 +  O(F^3) \, ,
\nonumber\\
\ddot{\cal G}_B\garg  
&=& \ddot G_B\garg +2i\dot G_B\garg eF
-4\Bigl(G_B\garg-{T\over 6}\Bigr){(eF)}^2+O(F^3) \, .
\nonumber\\
\end{eqnarray}
\noindent
These expansions are easily obtained from \eqref{app-GB} using the identity $\dot G_B^2\garg = 1-{4\over T}G_B\garg$.
The coefficients can be written in closed form to all orders in $F$, either in terms of Bernoulli polynomials of $\tau-\tau'$ \cite{18},
or in terms of Faulhaber polynomials of $\dot G_B\garg$ \cite{76}. 

\subsection{Matrix decomposition of the Green's function ${\cal G}_B$}

Finally, a matrix decomposition of ${\cal G}_B$ will be necessary. This can be achieved in a Lorentz invariant way \cite{40}, but
from a practical point of view it is simpler to work in a Lorentz frame well-adapted to the external field.
Here, we will be satisfied with treating (i) the case of a generic field and (ii) the purely magnetic field case;
see \cite{40} for the more special `crossed field' and `self-dual' cases. 
In all cases it will be useful to decompose ${\cal G}_B$ as

\bear
{\cal G}_B
&=&
{\cal S}_B
+
{\cal A}_B\,,
\label{decomposecalGB}
\ear\no
where ${\cal S}_B$ is the even part of ${\cal G}_B$ as a function of $F$, and ${\cal A}_B$ the odd one. 
For ${\cal S}_B$, the following trigonometric rewriting is often useful:

\bear
{\cal S}_B(\tau,\tau') -{ \cal S}_B(\tau,\tau)  = T \frac{\sin \bigl( \abs{u-u'}\cZ\bigr) \sin[(1-\abs{u-u'})\cZ]}{\cZ\sin \cZ}\,,
\label{RecalG}
\ear
where we have rescaled $\tau = Tu, \tau' = Tu'$. 

\subsubsection{The generic case}

For a generic constant field, both Maxwell invariants ${\bf B}^2 - {\bf E}^2$ and ${\bf E}\cdot{\bf B}$ are nonzero. 
By Lorentz invariance there then exists a Lorentz frame where the electric and magnetic field vectors both point along the $z$ - axis,
and by parity invariance we can assume that they both point along the {\it positive} $z$ - axis. In euclidean conventions, we then have

\begin{equation}
F =
\left(
\begin{array}{*{4}{c}}
0&B&0&0\\
-B&0&0&0\\
0&0&0&iE\\
0&0&-iE&0
\end{array}
\right),
\end{equation}
which suggests to introduce the following matrix base:

\begin{equation}
g_{\perp}\equiv
\left(
\begin{array}{*{4}{c}}
1&0&0&0\\
0&1&0&0\\
0&0&0&0\\
0&0&0&0
\end{array}
\right), \, 
g_{\parallel}\equiv
\left(
\begin{array}{*{4}{c}}
0&0&0&0\\
0&0&0&0\\
0&0&1&0\\
0&0&0&1
\end{array}
\right),\, 
r_{\perp} \equiv
\left(
\begin{array}{*{4}{c}}
0&1&0&0\\
-1&0&0&0\\
0&0&0&0\\
0&0&0&0
\end{array}
\right),\, 
r_{\parallel} \equiv
\left(
\begin{array}{*{4}{c}}
0&0&0&0\\
0&0&0&0\\
0&0&0&1\\
0&0&-1&0
\end{array}
\right)
\, .
\nonumber\\
\label{defBmatrices}
\end{equation}
Using this Lorentz frame and base, 
and defining

\bear
z_{\perp} \equiv eBT, \quad
z_{\parallel} \equiv ieET\,,
\label{defzpp}
\ear
the matrix functions ${\cal S}_B$ and ${\cal A}_B$ can be decomposed as \cite{40,41}

\bear
{\cal S}_{B12}^{\mu\nu}
&=&
-{T\over 2}
\sum_{\alpha ={\perp},{\parallel}}
{A_{B12}^{\alpha}\over z_{\alpha}}\,g_{\alpha}^{\mu\nu}\,,
\nonumber\\
{\cal A}_{B12}^{\mu\nu}
&=&
{iT\over 2}
\sum_{\alpha ={\perp},{\parallel}}
{S_{B12}^{\alpha}-\dot G_{B12}\over z_{\alpha}}
\,r_{\alpha}^{\mu\nu}\,,
\nonumber\\
\dot{\cal S}_{B12}^{\mu\nu} &=&
\sum_{\alpha ={\perp},{\parallel}}
S_{B12}^{\alpha}\,g_{\alpha}^{\mu\nu}\,,
\nonumber\\
\dot{\cal A}_{B12}^{\mu\nu} &=& 
-i
\sum_{\alpha ={\perp},{\parallel}}
A_{B12}^{\alpha}\,r_{\alpha}^{\mu\nu}\,,
\nonumber\\
\ddot{\cal S}_{B12}^{\mu\nu} &=& \ddot G_{B12}g^{\mu\nu}
-{2\over T}
\sum_{\alpha ={\perp},{\parallel}}
z_{\alpha}A_{B12}^{\alpha}\,g_{\alpha}^{\mu\nu}\,,
\nonumber\\
\ddot{\cal A}_{B12}^{\mu\nu} &=& 
{2i\over T}
\sum_{\alpha ={\perp},{\parallel}}
z_{\alpha}S_{B12}^{\alpha}\,r_{\alpha}^{\mu\nu}\,,
\nonumber\\
\label{app-decompcalSA}
\ear\no
with the following coefficient functions:

\bear
S_{B12}^{\alpha} &=&
{\sinh(z_{\alpha}\dot G_{B12})\over \sinh(z_{\alpha})} \,,
\nonumber\\
A_{B12}^{\alpha} &=&
{\cosh(z_{\alpha} \dot G_{B12})\over 
\sinh(z_{\alpha})}-{1\over z_{\alpha}}\,.
\nonumber\\
\label{app-AB}
\ear\no
(${\alpha} = \perp,\parallel$). 
In the worldline formalism,
these two scalar, dimensionless functions $S_B$ and $A_B$ 
are the basic building blocks of the integrands of one-loop amplitudes in a constant field
in scalar QED, as well as in scalar Einstein-Maxwell theory \cite{61,76}. 

\subsubsection{The magnetic case}

For easy reference, let us write down here also the explicit formulas for the case of a pure magnetic field,
with $\bf B$ pointing along the $z$ - axis:

\begin{eqnarray}
\bar{\cal G}_{B12}
&=&G_{B12}\,{g_{\parallel}}
-{T\over 2}{\Bigl(\cosh(z\dot G_{B12})-\cosh(z)\Bigr)
\over z\sinh(z)}
{g_{\perp}}\nonumber\\
&&\hspace{1.5cm}+{T\over{2z}}\biggl({\sinh(z\dot G_{B12})\over\sinh(z)}
-\dot G_{B12}\biggr)i{r_\perp}\,,\nonumber\\
\dot{\cal G}_{B12}
&=&\dot G_{B12}\,{g_{\parallel}}+{\sinh(z\dot G_{B12})\over\sinh(z)}
{g_{\perp}}
-\biggl({\cosh(z\dot G_{B12})\over \sinh(z)}-{1\over z}
\biggr)i{r_{\perp}}\,,\nonumber\\
\ddot{\cal G}_{B12}
&=& \ddot G_{B12}\,{g_{\parallel}}
+2\biggl(\delta_{12}-{z\cosh(z\dot G_{B12})\over T
\sinh(z)}\biggr){g_{\perp}}
+2{z\sinh(z\dot G_{B12})\over T\sinh(z)}i{r_{\perp}}\,.\nonumber\\
\label{GB(F)pureB}
\end{eqnarray}
\noindent
Here the ``bar'' on ${\cal G}_B$ indicates that its irrelevant coincidence limit has been subtracted. 
The DBC Green's function in the magnetic case can be written relatively compactly as \cite{mckshe}

\bear
\sdel(\tau,\tau^{\prime})
&=& \Delta\garg g_{\parallel}+
\frac{2eT}{z}\Big[\theta(\tau-\tau')\sin \frac{z(\tau-\tau')}{2eT}-\frac{\sin \frac{z\tau}{2eT}\sin \frac{z(T-\tau')}{2eT}}{\sin \frac{z}{2e}}\Big]\non
&&\times\Big[\cos\frac{z(\tau-\tau')}{2eT}g_\perp + \sin \frac{z(\tau-\tau')}{2eT}r_\perp \Big]\,,\non
\ear
where now $z= eBT$.


\end{document}